\documentclass[traditabstract,letter]{aa}

\usepackage{graphicx}
\usepackage{txfonts}
\usepackage{float}

\usepackage{natbib}
\bibpunct{(}{)}{;}{a}{}{,}

\usepackage{hyperref}
\hypersetup{colorlinks=true,
            citecolor=blue,
            linkcolor=blue}

\def\vv#1{\vec{#1}}
\def\hf{h_\mathrm{f}}
\def\sun{\odot}
\def\ear{\oplus}
\def\s{*}
\def\M{m_\s}
\def\rs{r_0} 
\def\Rs{R_\s}

\def\t{s}
\def\Rt{R_\t}
\def\Vt{V_\t}
\def\rhot{\rho_\t}
\def\R{R}
\def\X{X}
\def\Y{Y}
\def\Z{Z}
\def\r{r}
\def\x{x}
\def\y{y}
\def\z{z}
\def\ic{{i}}
\def\vA{{\cal A}}
\def\vI{{\cal I}}
\def\vS{{\cal S}}

\def\aquismall#1{{\small #1}}

\def \llabel#1{\label{#1}}

\begin{document}

\title{Transit light curve and inner structure of close-in planets}
\titlerunning{Transit light curves of close-in planets}

\author{
Alexandre C.M.~Correia\inst{1,2}
}

 
\institute{Departamento de F\'isica, I3N, Universidade de Aveiro, Campus de
Santiago, 3810-193 Aveiro, Portugal
  \and 
ASD, IMCCE-CNRS UMR8028,
Observatoire de Paris, UPMC,
77 Av. Denfert-Rochereau, 75014 Paris, France  
}

\date{Received ; accepted To be inserted later}

\abstract{Planets orbiting very close to their host stars have been found, some of them on the verge of tidal disruption.
The ellipsoidal shape of these planets can significantly differ from a sphere, which modifies the transit light curves. 
Here we present an easy method for taking the effect of the tidal bulge into account in the transit photometric observations.
We show that the differences in the light curve are greater than previously thought.
When detectable, these differences provide us an estimation of the fluid Love number, which is invaluable information on the internal structure of close-in planets.
We also derive a simple analytical expression to correct the bulk density of these bodies, that can be 20\% smaller than current estimates obtained assuming a spherical radius.
}


   \maketitle
%



\section{Introduction}

About half of the more than 1000 known transiting exoplanets have orbital periods less than ten days\footnote{http://exoplanet.eu/}.
These close-in planets undergo strong tidal effects raised by the parent star.
One consequence is that their spins and orbits evolve until an equilibrium configuration is reached, corresponding to coplanarity, circularity, and synchronous rotation \citep[e.g.,][]{Hut_1980, Correia_2009}.
Another consequence is that the shape of these planets differ from a spherical body, and it is approximated better by a triaxial ellipsoid \citep[e.g.,][]{Chandrasekhar_1987, Correia_Rodriguez_2013}.
The asymmetry in the mass distribution increases with the proximity to the star, and it is particularly pronounced near the Roche limit \citep[e.g.,][]{Ferraz-Mello_etal_2008, Burton_etal_2014}.

For simplicity, most observational works on transiting planets assume that its shape is spherical, so they determine an average radius.
However, 
if there is enough precision in the data, it is possible to spot the polar oblateness ($f_c$) 
signature 
in the transit light curve \citep{Seager_Hui_2002, Barnes_Fortney_2003, Ragozzine_Wolf_2009, Carter_Winn_2010a}, which gives us invaluable information on their internal structure. 
Previous studies have ignored the equatorial prolateness ($f_a$), 
which is actually more pronounced than the polar oblateness, 
since the long axis always points to the star, and thus should not be perceptible during the transit.
\citet{Leconte_etal_2011a} and \citet{Burton_etal_2014} have shown that failing to account for this distortion would result in a systematic underestimation of the planetary radius, hence of its bulk density.
In addition, for planets near the Roche limit, the projected ellipsoid also depends on the inclination to the line of sight and on the rotation angle, which most likely modifies the transit light curve.
Unfortunately, there is no comparison of transit light curves for ellipsoidal and spherical planets shown in the 
works by \citet{Leconte_etal_2011a} and \citet{Burton_etal_2014}.

Since a large number of transiting planets are being detected on the verge of tidal disruption \citep[e.g.,][]{Valsecchi_Rasio_2014}, it becomes important to understand the exact contribution of its shape to the photometric observations. 
The work by \citet{Leconte_etal_2011a} is very complete, but as pointed by \citet{Burton_etal_2014}, it requires complex internal structure models that are difficult to implement and to reconcile with observational parameters.
Therefore, \citet{Burton_etal_2014} compute the tidal deformation by assuming surfaces of constant gravitational equipotential for the planet that are solely based on observable parameters.
Nevertheless, this method is also not easy to implement, since it requires a numerical adjustment, such that the projected area of the model planet during the transit matches those given by the observations.
Moreover, we lose the information relative to the internal structure, which is an important complement to the density determination.


In this Letter, we propose a simple analytical model for computing the projected area of close-in planets at any point of its orbit, which is based in the equilibrium surface given by second-order Love numbers. 
We thus obtain the transit light curve for these planets, which can be used to compare directly with the observations, and infer their internal structure and density.


\section{Shape}

The shape (or the figure) of a planet is usually described well by a reference 
 ellipsoid (quadrupolar approximation).
The standard equation of a triaxial ellipsoid centered on the origin of a Cartesian coordinate system \aquismall{$(\X,\Y,\Z)$} and aligned with the axes is
\begin{equation}
\frac{\X^2}{a^2} + \frac{\Y^2}{b^2} + \frac{\Z^2}{c^2} = 1 \ , \llabel{140627b}
\end{equation}
where $a$, $b$, and $c$ are called the semi-principal axes. 
For the Earth and the gaseous planets in the solar system, we have $a \simeq b > c$ (oblate spheroids), but usually $a > b > c$ for the main satellites, where the long axis $a$ is directed to the central planet.

We let $\vv{\R} = (\X,\Y,\Z)$ be a generic point  at the surface of the ellipsoid.
Then, equation (\ref{140627b}) can be rewritten as
 \begin{equation}
\vv{\R}^T \vA_0 \, \vv{\R} = 1 \ ,
\quad \mathrm{with} \quad \vA_0 = \left[\begin{array}{ccc}  a^{-2} &  0 & 0 \\ 0 &  b^{-2} &0 \\ 0 &  0 & c^{-2} \end{array}\right] 
\ , \llabel{140716z}
\end{equation}
where 
$^T$ denotes the transpose.
Solving equation (\ref{140627b}) for $\Y^2$, the radial distance $\R = |\vv{\R}| $ can be expressed as 
\begin{equation}
\R^2 = 
\X^2 + \Y^2 +\Z^2 = b^2 \left( 1 + \alpha \, \X^2 + \gamma \, \Z^2  \right) \ , \llabel{140701a}
\end{equation}
where
\begin{equation}
\alpha = \left(\frac{1}{b^2} - \frac{1}{a^2}\right) = \frac{f_a \, (2+ f_a)}{b^2 \, (1+ f_a)^2} \ , \quad \mathrm{with} \quad f_a = \frac{a-b}{b}  \ , \llabel{140701b}
\end{equation}
and
\begin{equation}
 \gamma = \left(\frac{1}{b^2} - \frac{1}{c^2} \right) = \frac{f_c \, (2+ f_c)}{b^2 \, (1+ f_c)^2} \ , \quad \mathrm{with} \quad f_c = \frac{c-b}{b}  \ . \llabel{140701c}
\end{equation}
Assuming $ a \sim b \sim c $, we can neglect terms in $f_a^2$ and $f_c^2$ in the previous expressions, and thus
\begin{equation}
\R \approx b  \left( 1 + \frac{\alpha}{2} \X^2 + \frac{\gamma}{2} \Z^2  \right) \approx b + \frac{f_a}{b} \X^2 + \frac{f_c}{b} \Z^2 \ . \llabel{140701d}
\end{equation}

The mass distribution inside the planet is a result of the self gravity, but also of the body's deformation in response to any perturbing potential $V_p$.
A very convenient way to define this deformation is through the Love number approach \citep[e.g.,][]{Love_1911} 
in which the radial displacement $\Delta \R$  is proportional to the equipotential perturbing surface
\begin{equation}
\Delta \R = - \hf V_p / g \ , \llabel{131021a}
\end{equation}
where $g = G m / \R^2$ is the surface gravity, $G$ is the gravitational constant, $m$ is the mass of the planet, and $\hf$ is the fluid 
second Love number for radial displacement.
For a homogeneous sphere $\hf = 5/2$, but more generally, it can can be obtained from the Darwin-Radau equation \citep[e.g.,][]{Jeffreys_1976}
\begin{equation}
\frac{\vI}{m R^2} = \frac{2}{3} \left[ 1 - \frac{2}{5} \left( \frac{5}{\hf}-1\right)^{1/2} \right]
\ , \llabel{131021b}
\end{equation}
where $\vI$ is the mean moment of inertia, which depends on the internal mass differentiation.

Like the main satellites in the solar system, close-in planets deform under the action of the centrifugal and tidal potentials. 
The first results from the planet's rotation rate $\Omega$ about the $c$ axis, while the second results from the differential attraction of a mass element by the nearby star with mass $\M$.
For simplicity, we consider that the planet reached the final tidal equilibrium; that is, its orbit is circular with radius $\rs$, the spin axis is normal to the orbital plane (zero obliquity), and the rotation rate is synchronous with the orbital mean motion $n$, always pointing the long axis $a$ to the star \citep[e.g.,][]{Ferraz-Mello_etal_2008}.
Thus, on the planet's surface, the non-spherical contribution of the perturbing potential is given by \citep[e.g.,][]{Correia_Rodriguez_2013}
\begin{equation}
V_p  = \frac{1}{2} \Omega^2 \Z^2 - \frac{3 G \M}{2 \rs^3} \X^2  \ . \llabel{130529a}
\end{equation}

Replacing the above perturbing potential in expression (\ref{131021a}) and comparing with equation (\ref{140701d}) for the surface of the ellipsoid, it becomes straightforward that
\begin{equation}
\frac{f_a}{b}  =  \hf \frac{3 G \M}{2 g \rs^3}  \quad \Leftrightarrow \quad 
a  =  b \left( 1 + 3 q \right)  \ , \llabel{140701f}
\end{equation}
and
\begin{equation}
\frac{f_c}{b}  = - \frac{\hf}{2 g} \Omega^2  \quad \Leftrightarrow \quad 
c  =  b \left( 1 - q  \right)  \ , \llabel{140701g}
\end{equation}
with 
\begin{equation}
q = \frac{\hf}{2} \frac{\M}{m} \left(\frac{b}{\rs} \right)^3   \ , \llabel{140702a}
\end{equation}
where  $\Omega^2 = n^2 \approx G \M / \rs^3 $.
Since $q \propto \rs^{-3}$, the closer the planet is to the star, the greater is the difference between the ellipsoid semi-axes.
However, there is a maximum value for $q$, corresponding to the Roche limit, $\r_R$ \citep{Chandrasekhar_1987}
\begin{equation}
 \rs \gtrsim \r_R \equiv 2.46 \left( \frac{\M}{m} \right)^{1/3} b \quad \Rightarrow \quad q < q_{max} \approx \frac{\hf}{30} \ . \llabel{140702b}
\end{equation}
Adopting the maximum value possible for $\hf = 5/2 $ (Eq.\,\ref{131021b}) gives $ q_{max} \approx 1/12 \approx 0.08$, $a_{max} \approx 1.25 \, b$, and $c_{max} \approx 0.92 \, b $. 
Thus, the approximation used to obtain expression (\ref{140701d}) is still valid.

\section{Transit light curve}


We let  $(\x,\y,\z)$ be the standard Cartesian coordinate system for exoplanets, centered on the star, where $(\x,\y)$ are in the plane of the sky and $\z$ is along the line of sight.
In this frame, a generic point on the planet's surface $\vv{\r} = (\x,\y,\z)$ can be obtained as
\begin{equation}
\vv{\r} = \vv{\r}_0 + \vS \, \vv{R} \ , \llabel{140716a}
\end{equation}
where $\vv{\r}_0 =  (\x_0, \y_0, \z_0)$ is the position of the center of mass of the planet with respect to the star, and $\vS$ is a rotation matrix.
When the planet orbits the star in a synchronous circular orbit with zero obliquity, we have
\begin{equation}
\vS = \vS_x (\ic) \, \vS_z (\theta) = \left[\begin{array}{ccc}  \cos \theta & -\sin \theta & 0  \\ \sin \theta \cos \ic  & \cos \theta \cos \ic & -\sin \ic \\ \sin \theta \sin \ic &  \cos \theta \sin \ic & \cos \ic \end{array}\right] \ , \llabel{140716b}
\end{equation}
where $\theta$ is the rotation angle, $\ic$ is the inclination angle between the normal to the orbit  
and the line of sight, and $\vS_x$ and $\vS_z$ are the standard $3 \times 3$ 
rotation matrix about the $x$ and $z$ axis, respectively \citep[e.g.,][]{Murray_Correia_2010B}.
In this particular tidal equilibrium configuration, for a given time $t$, we have $\theta = n (t-t_0)$, and thus $\vS=\vS(t)$.
Moreover, the position of the planet in its orbit can also be easily obtained 
from $\vS$ as
\begin{equation}
\vv{\r}_0 = \vS \, \vv{R}_0 
= \rs \left[\begin{array}{c}  \cos \theta \\ \sin \theta \cos \ic \\ \sin \theta \sin \ic \end{array}\right]  \ ,
\quad \mathrm{where} \quad  
\vv{R}_0 = \left[\begin{array}{c}  \rs \\ 0 \\ 0 \end{array}\right] 
\ . \llabel{140722a}
\end{equation}

From expression (\ref{140716a}) we have
\begin{equation}
\vv{R} = \vS^{-1} (\vv{\r} - \vv{\r}_0) = \vS^T (\vv{\r} - \vv{\r}_0) \ , 
\quad \mathrm{and} \quad 
\vv{R}^T =  (\vv{\r} - \vv{\r}_0)^T \vS 
\ . \llabel{140716c}
\end{equation}
Replacing (\ref{140716c}) in equation (\ref{140716z}), we obtain the equation for the ellipsoidal surface of the planet in the new coordinate system
\begin{equation}
(\vv{\r}-\vv{\r}_0)^T \vA \, (\vv{\r}-\vv{\r}_0) = 1 \ , 
\quad \mathrm{with} \quad 
\vA = \vS \, \vA_0 \, \vS^T 
\ . \llabel{140627z}
\end{equation}

The projection of the ellipsoid in the ($\x,\y$) plane of the sky is obtained simply by setting $\z = \z_0$ in previous equation.
The result is an ellipse with general equation given by
\begin{equation}
A (\x-\x_0)^2 + B (\x-\x_0) (\y-\y_0) + C (\y-\y_0)^2 = 1 \ , \llabel{140717e}
\end{equation}
with
\begin{equation}
 A = \cos^2 \theta /a^2+\sin^2 \theta/b^2 \ , \llabel{140717a}
\end{equation}
\begin{equation}
B = (1/a^2-1/b^2) \sin 2 \theta \cos \ic  \ , \llabel{140717b}
\end{equation}
\begin{equation}
C = (\sin^2 \theta /a^2 + \cos^2 \theta/b^2) \cos^2 \ic + \sin^2\ic/c^2 \ . \llabel{140717c}
\end{equation}

\begin{figure*}
\begin{center}
\includegraphics[width=17.3cm]{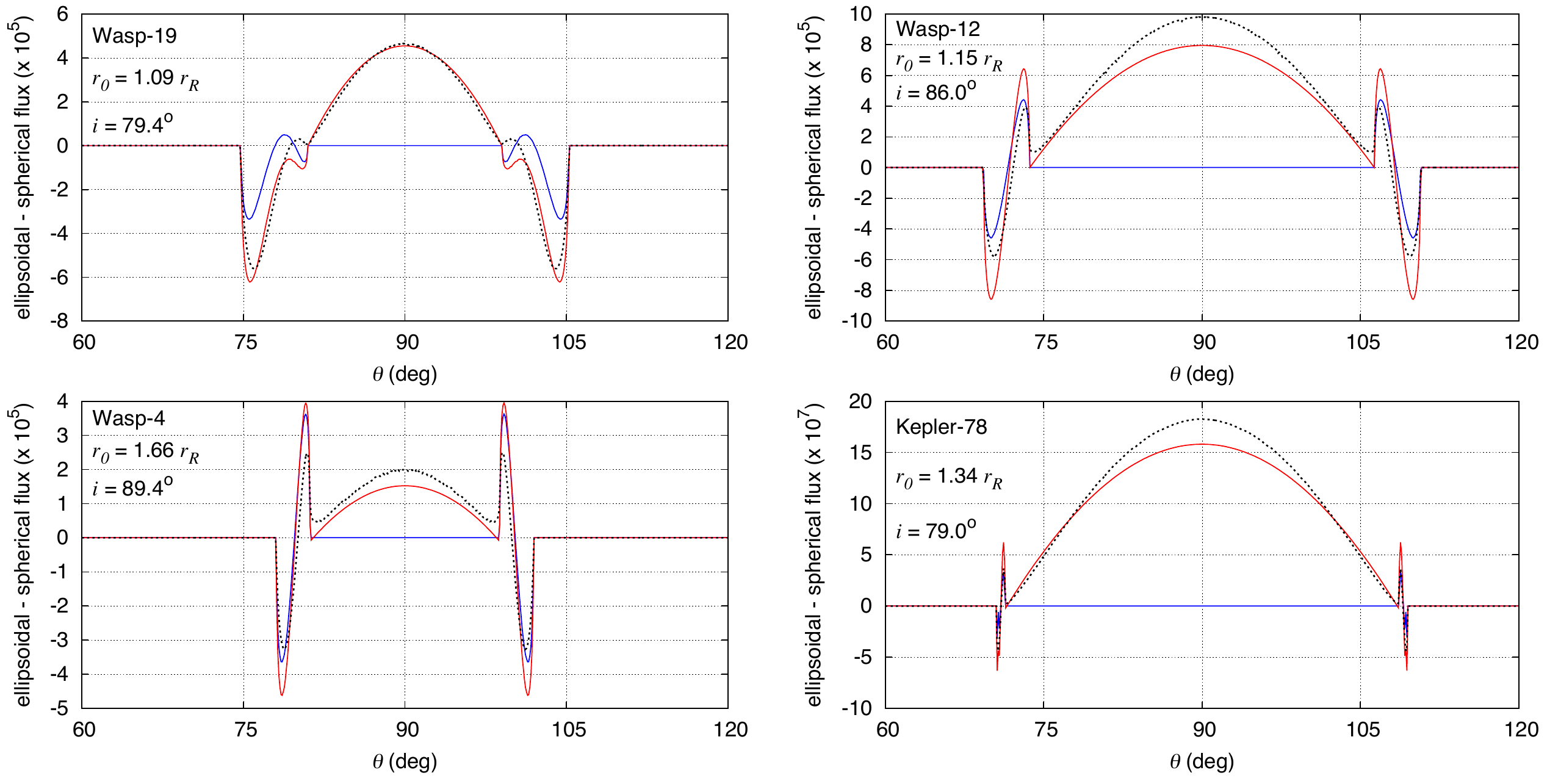} 
 \caption{Difference between the transit light curves of an ellipsoidal and a spherical planet, for some real close-in planets with different inclinations and distances to the star (Table~\ref{T1}).
The red line shows the difference obtained with our model (Eq.\,\ref{140717e}), while the blue line is obtained for an oblate planet with $a = b \ne c$, both with stellar uniform flux.
The dashed line also corresponds to the difference obtained with our model, but using a quadratic stellar limb-darkening correction $I(\mu)/I(1) = 1 - 0.4830  (1-\mu) - 0.2023 (1-\mu)^2$, which corresponds to a stellar temperature of $5500$~K \citep{Sing_2010}.
 \llabel{figreal}  }
\end{center}
\end{figure*}

In the new coordinate system, the diminishing of the stellar flux is then given by the overlap between the planetary ellipse (Eq.\,\ref{140717e}) and the stellar disk, defined by
\begin{equation}
\x^2 + \y^2 = \Rs^2
\ , \llabel{140717d}
\end{equation}
where we assumed a spherical shape for the star with radius $\Rs$.
There are many different methods of computing the overlap area.
If the stellar flux is uniform, one easy way is to compute the segment areas \citep[e.g.,][]{Eberly_2008, Hughes_Chraibi_2012}. 
In more realistic problems (models that include limb-darkening, stellar rotation, stellar activity, etc.), one can use Monte Carlo integrations \citep[e.g.,][]{Press_etal_1992, Carter_Winn_2010a}.


In Figure~\ref{figreal} we plot  the flux difference between the 
transit light curve of an ellipsoidal 
and a spherical planet for some real close-in planets, chosen with different inclinations and distances to the star (Table~\ref{T1}). 
The projected ellipsoid is obtained with our model (Eq.\,\ref{140717e}), while
the spherical radius is chosen such that its circumference area is equal to the projected ellipse area
just after the interior ingress (or just before the interior egress). 
We also show the flux difference for an oblate planet with $a = b \ne c$, in order to compare with previous studies \citep{Seager_Hui_2002, Barnes_Fortney_2003, 
Carter_Winn_2010a}.

There are essentially two main differences between the ellipsoidal and the spherical cases. 
One occurs at the ingress and egress phases, which corresponds to an oscillation in the flux difference.
This feature is mainly due to the polar oblateness, and therefore it was already identified in previous studies \citep[e.g.,][]{Seager_Hui_2002}.
The second feature, previously unnoticed, corresponds to a ``bump'' increase in the light curve difference during the whole transit, which is due to the rotation of the planet.
For instance, for $\ic=90^\circ$, 
the semi-axes of the projected ellipse are given by $A^{-1/2} = (\cos^2 \theta /a^2+\sin^2 \theta/b^2)^{-1/2}$ and  $C^{-1/2} = c$ (Eqs.\,\ref{140717a}-\ref{140717c});
that is, when $a \ne b$, the long semi-axis $A^{-1/2}$ is a function of the phase $\theta$.
Thus, the larger is the phase span during the transit, the higher the difference with respect to the spherical case. 
The maximum value is reached at the center of the transit ($\theta = 90^\circ$).

The effect of an ellipsoidal planet in the transit light curve is also maximized for edge-on orbits ($\ic = 90^\circ$).
For lower (or higher) inclinations, the planet spends less time in front of the star, so the ``bump'' increase due to the rotation becomes smaller.
Nevertheless, for almost grazing orbits we can still observe significant differences in the light curve, although the two main features described above are no longer completely individualized (see Fig.\,\ref{figjup}, for a Jupiter-like planet at the Roche limit).

\section{Inner structure determination}

\llabel{newsec}

 \begin{table*}
 \caption{Observational data (left side) and derived parameters (right side) for some close-in planets near the Roche limit 
 ($\rs < 3 \, \r_R $). 
All planets listed here were simultaneously detected by transit and radial-velocity techniques.
 \label{T1} }
 \begin{center}
 \begin{tabular}{l | c c c c c c c c | c c c c c c c}
 \hline\hline
              &  $\rs$ & $\r_R$ &$\ic$ & $\M$ & $\Rs$ & $m$ & $\Rt$ & $\rhot$ & $\hf$ & $q$ & $a$ & $b$ & $c$ & $\rho$ & $\Delta \rho$ \\ 
Planet & ($\r_R$) & ($R_\sun$) & (deg) & ($M_\sun$) & ($R_\sun$) & ($M_\ear$) & ($R_\ear$)  & (g/cm$^{3}$) & & (\%) & ($R_\ear$) & ($R_\ear$) & ($R_\ear$) & (g/cm$^{3}$) & (\%) \\ 
 \hline
WASP-19b\,$^{1}$  & 1.09  & 3.27 & 79.4 &  0.97 & 0.99 & 371. & 15.2 & 0.58 & 1.5 & 3.92 & 17.3 & 15.4 & 14.8 & 0.51 & 12.9 \\ 
WASP-12b\,$^{2}$ & 1.15  & 4.30 & 86.0 &  1.35 & 1.60 & 446. & 19.0 & 0.36 & 1.5 & 3.34 & 21.3 & 19.3 & 18.7 & 0.31 & 11.6 \\ 
WASP-103b\,$^{3}$  & 1.19  & 3.59 & 86.3 &  1.22 & 1.44 & 473. & 16.8 & 0.55 & 1.5 & 2.99 & 18.5 & 17.0 & 16.5 & 0.50 & 10.4 \\ 
Kepler-78b\,$^{4}$ & 1.34  & 1.48 & 79.0 &  0.83 & 0.74 & 1.69 & 1.20 & 5.37 & 2.0 & 2.79 & 1.32 & 1.21 & 1.18 & 4.88 & 9.2 \\ 
WASP-52b\,$^{5}$ & 1.48  & 3.94 & 85.4 &  0.87 & 0.79 & 146. & 13.9 & 0.30 & 1.5 & 1.56 & 14.7 & 14.0 & 13.8 & 0.28 & 5.4 \\ 
CoRoT-1b\,$^{6}$ & 1.49  & 3.64 & 85.1 &  0.95 & 1.11 & 327. & 16.3 & 0.41 & 1.5 & 1.51 & 17.2 & 16.5 & 16.2 & 0.39 & 5.2 \\ 
OGLE-TR-56b\,$^{7}$ & 1.54 & 3.32 & 73.7 &  1.23 & 1.36 & 442. & 15.1 & 0.70 & 1.5 & 1.37 & 15.8 & 15.1 & 15.0 & 0.68 & 4.2 \\ 
WASP-78b\,$^{8}$ & 1.60  & 4.87 & 83.2 &  1.33 & 2.20 & 283. & 18.6 & 0.24 & 1.5 & 1.25 & 19.4 & 18.7 & 18.5 & 0.23 & 4.3 \\ 
WASP-48b\,$^{9}$ & 1.65  & 4.47 & 80.1 &  1.19 & 1.75 & 311. & 18.3 & 0.28 & 1.5 & 1.11 & 19.0 & 18.4 & 18.2 & 0.27 & 3.7 \\ 
WASP-4b\,$^{10}$ & 1.66  & 2.91 & 89.4 &  0.85 & 0.87 & 384. & 14.3 & 0.72 & 1.5 & 1.09 & 14.8 & 14.4 & 14.2 & 0.70 & 3.8 \\ 
HAT-P-23b\,$^{11}$ & 1.78  & 2.80 & 85.1 &  1.13 & 1.20 & 664. & 15.0 & 1.08 & 1.5 & 0.89 & 15.5 & 15.1 & 14.9 & 1.05 & 3.1 \\ 
WASP-43b\,$^{12}$  & 1.78  & 1.84 & 82.3 &  0.72 & 0.67 & 646. & 11.4 & 2.43 & 1.5 & 0.88 & 11.7 & 11.4 & 11.3 & 2.35 & 3.0 \\ 
55\,Cnc\,e\,$^{13}$  & 2.00  & 1.66 & 82.5 &  0.91 & 0.94 & 7.81 & 2.17 & 4.18 & 2.0 & 0.83 & 2.23 & 2.18 & 2.16 & 4.06 & 2.8 \\ 
WASP-18b\,$^{14}$ & 2.29  & 1.90 & 84.8 &  1.24 & 1.15 & 3235. & 16.7 & 3.78 & 1.5 & 0.42 & 17.0 & 16.8 & 16.7 & 3.73 & 1.4 \\ 
Kepler-10b\,$^{15}$  & 2.43  & 1.49 & 84.8 &  0.91 & 1.07 & 3.33 & 1.47 & 5.76 & 2.0 & 0.47 & 1.49 & 1.47 & 1.46 & 5.67 & 1.6 \\ 
CoRoT-7b\,$^{16}$  & 3.01  & 1.22 & 80.1 &  0.91 & 0.82 & 7.42 & 1.56 & 10.3 & 2.0 & 0.25 & 1.59 & 1.58 & 1.58 & 10.3 & 0.8 \\ 
  \hline
 \end{tabular}
 \end{center}
 Refs:
 $^{1}$\citet{Hellier_etal_2011}; 
 $^{2}$\citet{Chan_etal_2011}; 
 $^{3}$\citet{Gillon_etal_2014}; 
 $^{4}$\citet{Howard_etal_2013}; 
 $^{5}$\citet{Hebrard_etal_2013a};
 $^{6}$\citet{Barge_etal_2008};
 $^{7}$\citet{Adams_etal_2011}; 
 $^{8}$\citet{Smalley_etal_2012};
 $^{9}$\citet{Enoch_etal_2011};
 $^{10}$\citet{Gillon_etal_2009};
 $^{11}$\citet{Bakos_etal_2011};
 $^{12}$\citet{Gillon_etal_2012a};
 $^{13}$\citet{Gillon_etal_2012b}; 
 $^{14}$\citet{Southworth_etal_2009}; 
 $^{15}$\citet{Dumusque_etal_2014}; 
 $^{16}$\citet{Hatzes_etal_2011}. 
\end{table*}

For those systems where the differences in the light curve can be detected, one can use an ellipsoid instead of a sphere to fit the photometric observations (Eq.\,\ref{140717e}).
In addition to the equatorial radius $b$, the projected ellipsoid only depends on 
$q$ (Eqs.\,\ref{140701f}$-$\ref{140702a}).
Then, when we adjust our model to the observational data, $q$ is the only supplementary parameter to fit, which accounts for all the observed differences in the transit light curve.
In the expression of $q$ (Eq.\,\ref{140702a}), all parameters but $\hf$ are also known from the observational data.
It is thus possible to obtain an observational estimation for $\hf$, which represents an important additional constraint for the inner structure differentiation (Eq.\,\ref{131021b}).

\citet{Seager_Hui_2002} originally showed that the oblateness of transiting exoplanets can be constrained from the variations in the light curve during the ingress and egress phases.
However, \citet{Barnes_Fortney_2003} noticed that these oscillations are short in time and therefore difficult to observe, although some relatively weak constraints on the Love number were derived for the system HD\,189733 \citep{Carter_Winn_2010a}. 

In Figure~\ref{figreal} we observe that 
the ``bump'' increase due to the rotation of the prolate planet is present during the whole transit, not only at ingress and egress phases.
As a consequence, the flux differences can be significantly larger than previously thought, increasing our chances of observing the shape of close-in planets.
For giant planets near the Roche limit, the flux differences can reach $10^{-4}$, which is above the intrinsic stellar photometric variability on transit timescales, estimated to be $\sim 10^{-5}$ \citep{Borucki_etal_1997}.
For longer distances to the star, 
the flux difference can drop below the $10^{-5}$ threshold, so the shape is 
more difficult to determine.
For rocky planets, the flux difference is also smaller than $10^{-5}$.
However, since the Roche limit is closer to the star for these planets, 
the ``bump'' becomes more prominent, increasing again the chances of detection.

In Figure~\ref{figreal} we also observe that the flux difference is about  
the same for a stellar uniform flux, or when limb-darkening is considered.
The reason is that the differences in the shape of the planet are so tiny with respect to the size of the star that locally the flux can be considered approximately uniform.
Therefore, for realistic light curves obtained with non-uniform stellar flux, one can still obtain the correction introduced by the axial asymmetry using a uniform flux model, which is much easier to compute.

\section{Density determination}

Transiting systems provide a radius $\Rt$ for the planet, hence a bulk density, when coupled with
the planetary mass determined from radial-velocity measurements. 
Most studies assume a spherical planet in order to work out the volume $\Vt = 4 \pi \Rt^3 / 3 $, thus providing a spherical bulk density $\rhot = m / \Vt$. 
However, for close-in planets, the true volume of the ellipsoid is given by $V = 4 \pi a b c / 3$, which gives for the bulk density (Eqs.\,\ref{140701f}, \ref{140701g})
\begin{equation}
\rho = \frac{3 m}{4 \pi a b c} \approx \frac{3 m}{4 \pi b^3}  \left( 1 - 2 q \right)
\ . \llabel{140727a}
\end{equation}

When there is enough precision in the data, $b$ and $q$ can be directly adjusted from the transit light curve, instead of $\Rt$ (section~\ref{newsec}).
For low-mass planets such as Kepler-78b, it may be difficult to spot the tiny differences in the light curve with respect to a spherical planet (Fig.\,\ref{figreal}).
In those cases, one can use the spherical model, where $\Rt$ is related to the projected ellipsoid (Eq.\,\ref{140717e}) through 
\begin{equation}
\pi \Rt^2 
=  \frac{2\pi}{\sqrt{4 A C - B^2}} 
\ . \llabel{140727b}
\end{equation}
At the center of the transit ($\theta = 90^\circ $), we thus have (Eqs.\,\ref{140717a}-\ref{140717c})
\begin{equation}
b \approx \Rt \left(1 + q/2 - 2 q \cos^2 \ic \right)
\ , \llabel{140727c}
\end{equation}
which gives for the bulk density (Eq.\,\ref{140727a})
\begin{equation}
\rho 
\approx \rhot \left(1 -  7q/2 + 6 q \cos^2 \ic \right)
\ . \llabel{140727d}
\end{equation}
Here, $q$ cannot be adjusted from the observations, but it can be estimated from expression (\ref{140702a}) using $b \approx \Rt$, and adopting the solar system data,
$\hf = 1.5$ for gaseous planets, and $\hf = 2$ for rocky planets \citep{Yoder_1995cnt}.

In Table~\ref{T1} we compute the relative change in the bulk density, $\Delta \rho = 1 - \rho/\rhot $, for some close-in planets near the Roche limit ($\rs < 3 \, \r_R $).
These corrections agree with those obtained numerically by \citet{Burton_etal_2014}.
The small differences observed between the two methods result from the assumption of different inner structure models. 
Indeed, an exact match could be obtained using slightly different $\hf$ values in our model.

In Figure~\ref{figA} we show the relative change in the bulk density as a function of the distance to the star $\rs$.
The largest possible correction in the density is $\rho_{max} \approx  0.7 \, \rhot  $, obtained from $q_{max}$ (Eq.\,\ref{140702b}) with $\ic = 90^\circ$, which is 30\% smaller than the spherical value.
However, for realistic cases of planets  near the Roche limit ($\rs \approx \r_R$), we get values lower than 20\% for rocky planets and lower than 15\% for gaseous planets.
As in \citet{Burton_etal_2014}, we also observe that the relative change in the density decreases rapidly with the distance to the star.
This correction is less than 5\% for planets with $\rs > 1.5 \, \r_R$, which is equivalent to the measurement error on the currently published bulk densities \citep[e.g.,][]{Hellier_etal_2011}.
Therefore, with the current observational precision, a correction in the density determination is only justified for planets that are close to the Roche limit.

\begin{figure}
\begin{center}
\includegraphics[width=\columnwidth]{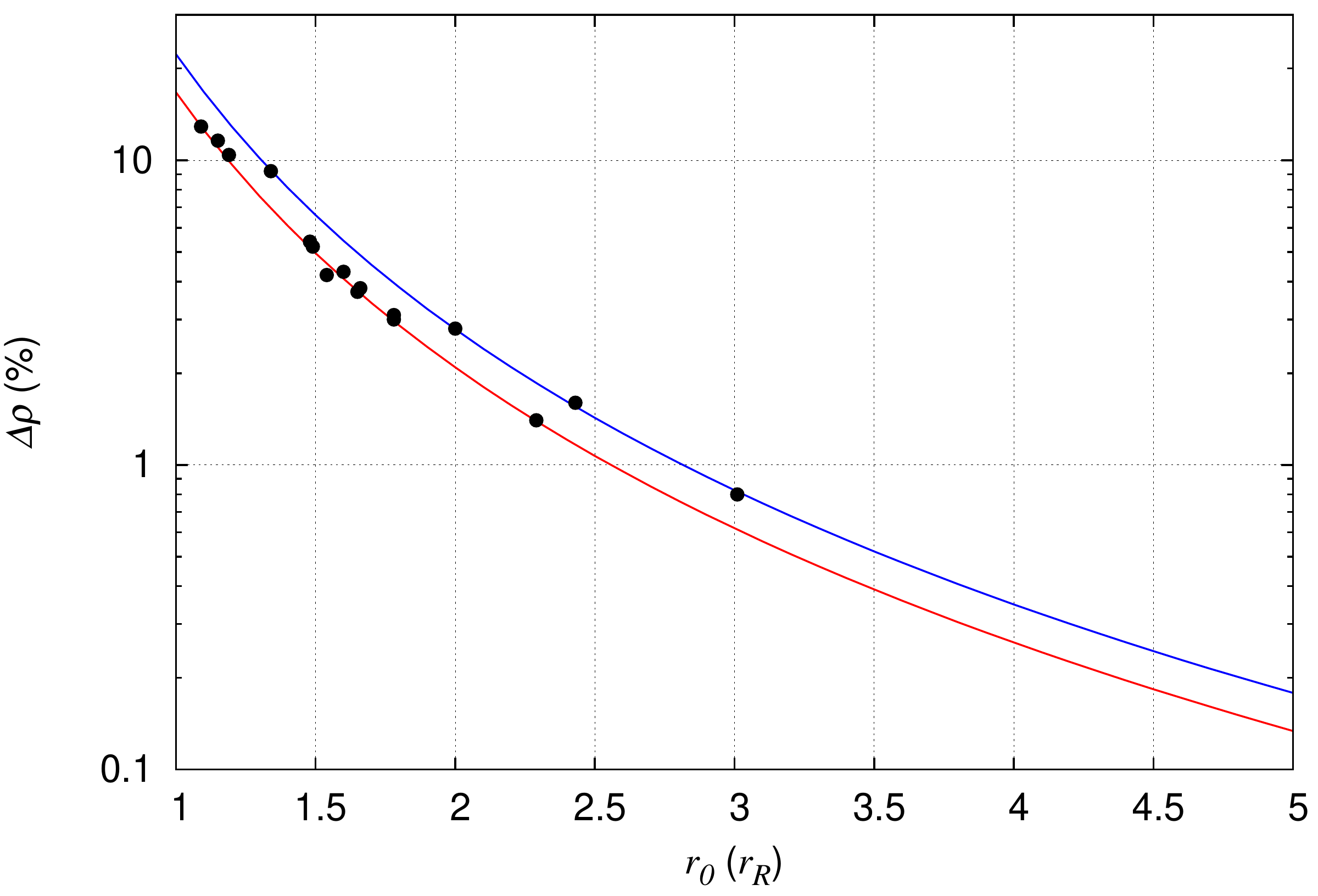} aqui
 \caption{Relative change in the bulk density of close-in planets, $\Delta \rho = 1 - \rho/\rhot $, as a function of the distance to the star $\rs$. The red curve is relative to gaseous planets ($\hf=1.5$), while the blue one is relative to rocky planets ($\hf=2.0$). Both curves are obtained using expression (\ref{140727d}) with $\ic = 80^\circ$. The dots represent the planets listed in Table~\ref{T1}. For $\rs > 1.5 \, r_R$, the change is negligible when compared to the present observational errors in the density determination.  \llabel{figA}  }
\end{center}
\end{figure}

\section{Conclusions}

We have presented here a simple model for correcting the transit light curve of close-in planets, which is easy to implement and which allows us to obtain important constraints for the internal structure of these planets.
On one hand, it provides a better determination for the bulk density (Eq.\,\ref{140727d}), which is lower than previously assumed (Fig.\,\ref{figA}).
For instance, it  has been announced that Kepler-78b has an Earth-like density \citep{Howard_etal_2013, Pepe_etal_2013}, but actually its real density is probably smaller than 5\,g/cm$^3$  (Table\,\ref{T1}).
On the other hand, information on the internal structure differentiation can be obtained by determining the fluid Love number $\hf$ (Eq.\,\ref{131021b}), even when the correction in the density is negligible.
If there is enough precision in the transit light curve ($< 10^{-4}$), one can adjust an ellipsoid instead of a sphere to the data (Fig.\,\ref{figreal}).
Then, in addition to the equatorial radius $b$, one can also quantify the planet asymmetry $q$, and thus obtain an observational estimation for $\hf$ (Eqs.\,\ref{140701f}-\ref{140702a}).

There are other methods that allow indirect determination of the Love number \citep[e.g.,][]{Batygin_etal_2009b, Ragozzine_Wolf_2009, Mardling_2010}, but they require that the global dynamics of the system is known (the presence of planetary companions, precise eccentricities and inclinations, etc.), which is very unlikely to achieve at present.
Our method does not require any additional knowledge on the system, apart from the variations in the transit light curve. 
It thus provides a direct determination of the Love number, so is therefore much more reliable.

In our model we have assumed that the planet is in a tidal final equilibrium configuration.
However, our method can be generalized to planets in eccentric orbits, with any rotation rate and non-zero obliquity.
For that purpose, a point at the surface of the planet (Eq.\,\ref{140716a}) is given by a new  rotation matrix
 \begin{equation}
\vS = \vS_x (\ic) \, \vS_z (\varphi) \, \vS_x (\varepsilon) \, \vS_z (\theta)  \ , \llabel{140727x}
\end{equation}
where $\varepsilon$ is the obliquity (the angle between the equator and the orbital plane), and $\varphi$ is the precession angle.
Also, we have now $\theta = \Omega (t-t_0)$, since $\Omega \ne n$.
Therefore, the position of the planet on its orbit (Eq.\,\ref{140722a}) also needs to be corrected through 
\begin{equation}
\vv{\r}_0 = \vS_x (\ic) \, \vS_z (\omega + v) \, \vv{R}_0 \ , \llabel{140727z}
\end{equation}
where $\omega$ is the argument of the pericentre and $v$ is the true anomaly.
For elliptical orbits with semi-major axis $a_0$ and eccentricity $e_0$, we also have $\rs = a_0 (1-e_0^2) / (1 + e_0 \cos v) $.

Finally, our model can also be used to correct the light curve of eclipsing close binary stars, where the secondary replaces the planet. For the primary, instead of using equation (\ref{140717d}), we need to use an analog to equation (\ref{140717e}), where $a$, $b$ and $c$ are replaced by the primary semi-axes and $\vv{r}_0 = 0$.
However, in this case the approximation for the disturbing potential (Eq.\,\ref{130529a}) may require additional corrective terms \citep[e.g.,][]{Budaj_2011, Burton_etal_2014}.

\begin{acknowledgements}
The author acknowledges support from Funda\c{c}\~ao para a Ci\^encia e a Tecnologia, Portugal (PEst-C/CTM/LA0025/2011), and from the PICS05998 France-Portugal program.
\end{acknowledgements}

\bibliographystyle{aa}
\bibliography{correia}

\appendix
\section{Additional figure}

\llabel{figuressec}

\begin{figure*}
\begin{center}
\includegraphics[width=\textwidth]{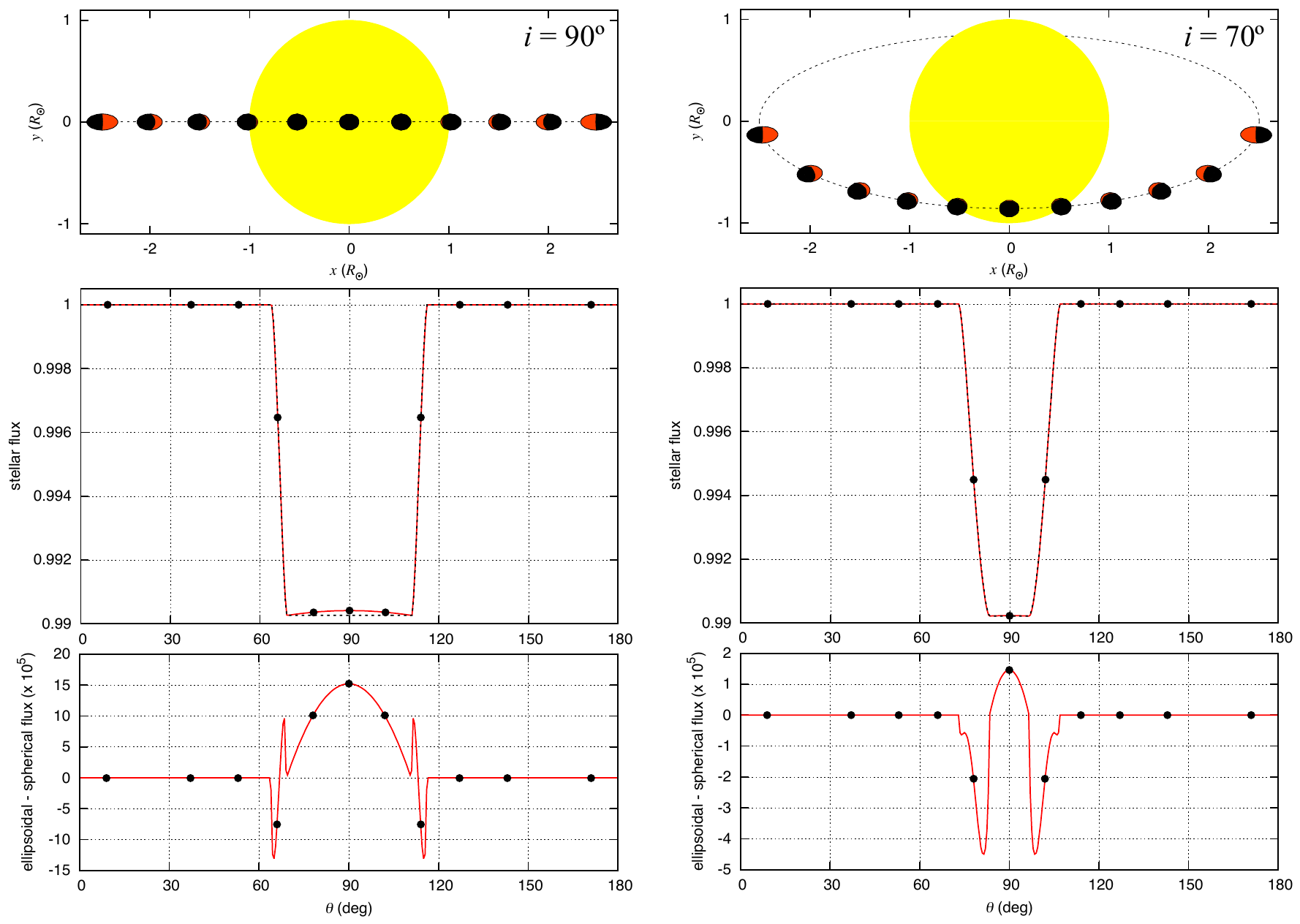} 
 \caption{Transit light curves for a Jupiter-like planet orbiting at the Roche limit ($\rs = \r_R = 2.51 \, R_\sun$) of a Sun-like star with uniform flux (in red).  
For comparison, we also show the light curve of a spherical planet (dashed line), and the difference between the two curves (bottom figure). The dots correspond to the relative positions shown in the top figure. 
The shape of the planet in the top figure is obtained with an exaggerated $\hf = 6$, so that we can better spot the modifications in the projected shape, but the transit light curves are obtained with the realistic Jupiter's value $\hf = 1.5$ \citep{Yoder_1995cnt}.
 \llabel{figjup}  }
\end{center}
\end{figure*}

\end{document}